\documentclass[pra,aps,twocolumn,superscriptaddress]{revtex4}
\usepackage{graphicx}
\usepackage{dcolumn}
\usepackage{amssymb}
\newcommand{\beqa}{\begin{eqnarray}}
\newcommand{\eeqa}{\end{eqnarray}}
\newcommand{\beq}{\begin{equation}}
\newcommand{\eeq}{\end{equation}}
\newcommand{\env}{\mathcal{E}}
\newcommand{\sys}{\mathcal{S}}
\newcommand{\info}{\mathcal{I}}
\newcommand{\entro}{\mathcal{H}}
\newcommand{\ent}{E}

\newcommand{\ket}[1]{| #1\rangle}

\begin{document}
\title{Redundancy of classical and quantum correlations during decoherence}

\author{Juan Pablo \surname{Paz}}
\affiliation{Departamento de F\'{\i}sica Juan Jos\'e Giambiagi, 
FCEyN, UBA, Pabell\'on 1, Ciudad Universitaria, 1428 Buenos Aires,
 Argentina}

\author{Augusto J. \surname{Roncaglia}}
\affiliation{Departamento de F\'{\i}sica Juan Jos\'e Giambiagi, 
FCEyN, UBA, Pabell\'on 1, Ciudad Universitaria, 1428 Buenos Aires,
 Argentina}

\begin{abstract}
We analyze the time dependence of entanglement and total correlations 
between a system and fractions of its environment in the course of 
decoherence. For the quantum Brownian motion model we show that the 
entanglement and total correlations have rather different dependence on 
the size of the environmental fraction. Redundancy manifests differently 
in both types of correlations and can be related with 
induced--classicality. To study this we introduce a new measure of 
redundancy and compare it with the existing one. 
\end{abstract}

\maketitle

\section{Introduction}

Decoherence is a physical process taking place when a system interacts with its environment. 
This process is crucial to understand the origin of the classical domain from a fundamentally 
quantum substrate \cite{Deco}. Absence of macroscopic
superpositions is one of the essential features of the classical realm which is explained by 
decoherence as a consequence of an effective super-selection rule that prevents the stable 
existence of the vast majority of states in the Hilbert space. In
recent years, important developments enabled us to understand the origin of another defining
property of classical systems: the fact that they exist in an objective state, a state of 
their own. Zurek and co-workers made this idea precise by noticing that the emergence of 
classicality is connected with the existence of redundant records of the state of the system
imprinted in the environment. Redundancy turns out to be the key notion to define objectivity 
and classicality: A system behaves classically when different fractions of any environment 
correlate with it in the same way (i.e. when correlations become redundant). Such consensus 
about the state of the system characterizes the classical realm and enables a definition of objectivity. 
In their pioneering work, Zurek and co-workers \cite{ZurekNat,Ollivier,Blume,Blume08,Zwolak09} 
analyzed the emergence of redundancy in the total
correlations between the system and the environment which can be measured by using information theoretic tools such as
mutual information. Here, we will examine how does redundancy manifests in purely quantum correlations (entanglement)
established between the system and fractions of the environment in the course of decoherence. We will present a definition
of redundancy based on the entanglement and show that the way in which redundancy manifests in total and quantum 
correlations is rather different.  

We will focus on the paradigmatic quantum Brownian motion (QBM) model \cite{Caldeira,HPZ}
where a particle $\sys$ interacts with an environment $\env$ that can be split into a 
fraction $\env_f$ and its complement $\env_{1-f}$ ($f$ parameterizes the size of the 
fraction $\env_f$ with respect to $\env$ ranging between $0$ and $1$).
Total correlations between $\sys$ and $\env_f$ can be measured with the 
mutual information $\info(\sys,\env_f)$ (computed from the reduced state obtained after
tracing out the complement $\env_{1-f}$). On the other hand, quantum correlations can be quantified 
with an entanglement measure between $\sys$ and $\env_f$, denoted as $\ent^{(f)}$. Here we will 
study both $\info(\sys,\env_f)$ and $\ent^{(f)}$ as a function of time and $f$, for random 
choices of the components of $\env_f$ out of the total $\env$. 

The paper is organized as follows: In Sec. \ref{sec:correl} we present 
the model and describe the temporal evolution of the correlations between the  
system and the different portions of the environment. We consider two types of  splittings of the environment in different fractions. First we analyze a splitting into bands oscillators (characterized by
their frequencies) and then we analyze grouping such bands into random fractions (characterized by their size). 
In Sec. \ref{sec:analyt} we derive analytical results for the mutual information and entanglement in the
non-dissipative regime. In Sec. \ref{sec:red} we study the evolution of redundancy of total correlations and
quantum correlations. In Sec. \ref{sec:conc} we summarize our results.

\section{Dynamics of correlations in quantum Brownian motion}
\label{sec:correl}

In the QBM model a central particle $\sys$ is coupled to an environment $\env$  
formed by harmonic oscillators $q_n$ ($n=1,..,N$). The total Hamiltonian is $H=H_\sys+H_{\sys\env}+H_\env$ where  
$H_\sys=p^2/2m+m \omega_\sys^2 x^2/2$, $H_\env=\sum_{n}\left(\frac{\pi_n^2}{2m_n}+\frac{m_n}{2} w_n^2 q_n^2\right)$ 
and $H_{\sys\env}=x\sum_{n}c_n q_n$ (we use $\hbar=1$ throughout). The spectral density $J(\omega)=\sum_{n}
{c_n^2}\delta(\omega-\omega_n)/{2m_n \omega_n}$
determines the effect of $\env$ on $\sys$. We consider a realistic family of environments where 
$J(\omega)=2 m\gamma_0\omega \left({\omega /\Lambda}\right)^{n-1}\theta(\Lambda-\omega)/\pi$ 
that includes Ohmic ($n=1$), super-Ohmic, ($n>1$) and sub-Ohmic ($n<1$) members 
($\Lambda$ is a high frequency cutoff, $\gamma_0$ a coupling strength, and $\theta(x)$ the Heaviside step function). 
We assume that the initial state of $\env$ is the ground state of $H_\env$ and that system-environment correlations vanish
initially, when the interaction is switch on correlations are created between system and environment \cite{Eisert}.
Quantum mutual information (MI) $\info(\sys,\env_f)$ \cite{MI,QMI} is defined from the von Neumann entropies of the
corresponding states:  $\info(\sys,\env_f)=\entro(\sys)+\entro(\env_f) -\entro(\sys,\env_f)$ 
(here $\entro(\sys)=-{\rm Tr}\left[\rho_\sys\ln\rho_\sys\right]$, etc.). 
The reduced state $\rho_{\{\sys,\env_f\}}$ is obtained after tracing out the complementary environment $\env_{1-f}$. We
restrict to initial Gaussian states which, due to linearity, remain Gaussian for
all times and can be described efficiently by a covariance matrix. In such case analytic expressions for
$\info(\sys,\env_f)$ can be obtained. In order to quantify the quantum correlations, we will use the multipartite
logarithmic negativity \cite{MulEnt} as a measure of entanglement between $\sys$ and $\env_f$. This is defined as the
maximum between zero and $-\sum_{i:\nu_i<1/2}\log(2\tilde\nu_i)$, where $\tilde\nu_i$ are the
symplectic eigenvalues of the partially transposed covariance matrix corresponding to the Gaussian state
$\rho_{\{\sys,\env_f\}}$. 

We present first numerical results and later we show analytic expressions 
that reproduce them with high accuracy. To do numerics we use a discrete version of the model 
\cite{Blume08,PazRoncaglia} with $600$ modes (we checked that our results are stable as this number is increased) and coupling strengths $c_n^2$ proportional to the spectral density. We set all masses to unity and take $\gamma_0=0.1$, the renormalized frequency $\Omega_\sys=3$, 
the cutoff frequency $\Lambda=20$ for dissipative environments ($n=1,1/2$) 
and we use a high cutoff, $\Lambda=300$, for the non-dissipative one ($n=3$). 
The initial state of the system is a pure Gaussian squeezed state
with a squeezing parameter $r=\ln(m\Omega_\sys \Delta x/\Delta p)=-5$ (unless it is specified) and $\Delta x\Delta p =1/2$.

\begin{figure}[ht]
\centering
\includegraphics[width=9cm]{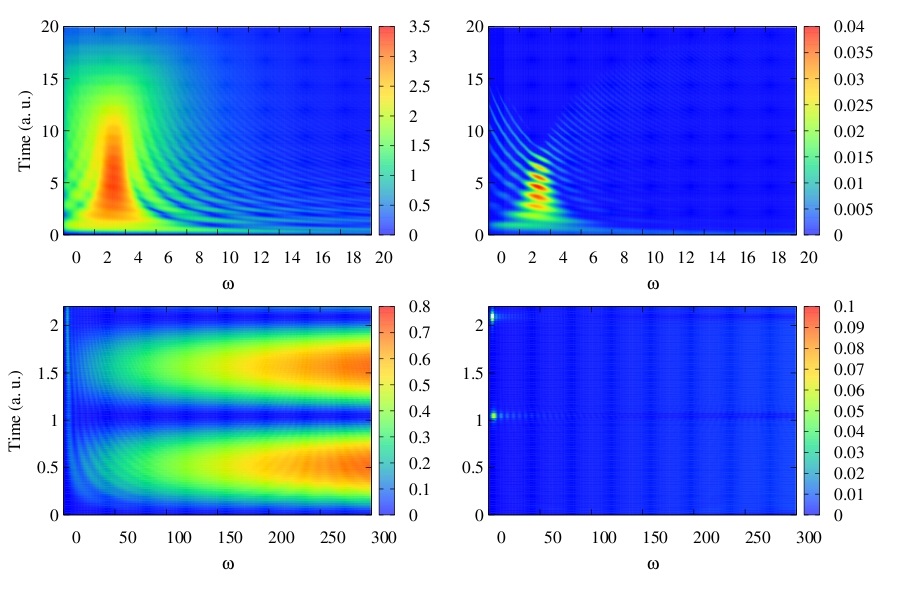}
\caption{Mutual Information (left) and logarithmic negativity (right) between a Brownian particle and bands of 
oscillators of the environment of frequency $\omega$. The initial state of the system is squeezed in position ($r=-5$). 
Plots at the top correspond to a sub-Ohmic environment while the ones at the bottom correspond to the super-Ohmic case. 
Ohmic environment behaves like sub-Ohmic one. } 
\label{fig:freq}
\end{figure}

\textit{Correlations with bands of oscillators:} Mutual information and entanglement between 
the system and environmental bands of frequency $\omega$ 
are shown in Fig. \ref{fig:freq}. MI develops fast and for short times concentrates at high frequencies.
For dissipative environments (sub-Ohmic case, $n=1/2$) at later times the resonant band is dominant. After a few relaxation times, 
the resonant peak is washed out and all oscillators in the environment tend to play a similar role. 
The entanglement has a similar behavior except for the super-Ohmic case. The resonant band of the environment is dominant for
entanglement and is also washed out by dissipation (contrary to the case of mutual information, entanglement with non-resonant
oscillators seems to be always negligible). One of our findings is that the super-Ohmic environment displays 
recoherence effects which manifest as oscillatory behavior both for the mutual information and entanglement. 
In this case, entanglement only develops with resonant oscillators.
Thus, the super-Ohmic environment does not induce truly dissipative effects but is responsible 
for a renormalization (dressing) of the system as well as for coherent oscillations which
will also be visible in our forthcoming studies. It is worth pointing out that the existence
of a perfect reversion for the super-Ohmic environment only takes place for some initial sates
of the system. If instead of the above state we choose one with a large momentum squeezing
complete recoherence is lost (see below). This was already noticed in \cite{PazHabZur93} and can be attributed to a process taking place at ultra-short timescales, as will be evident below. 

\begin{figure}[ht]
\centering
\includegraphics[width=9cm]{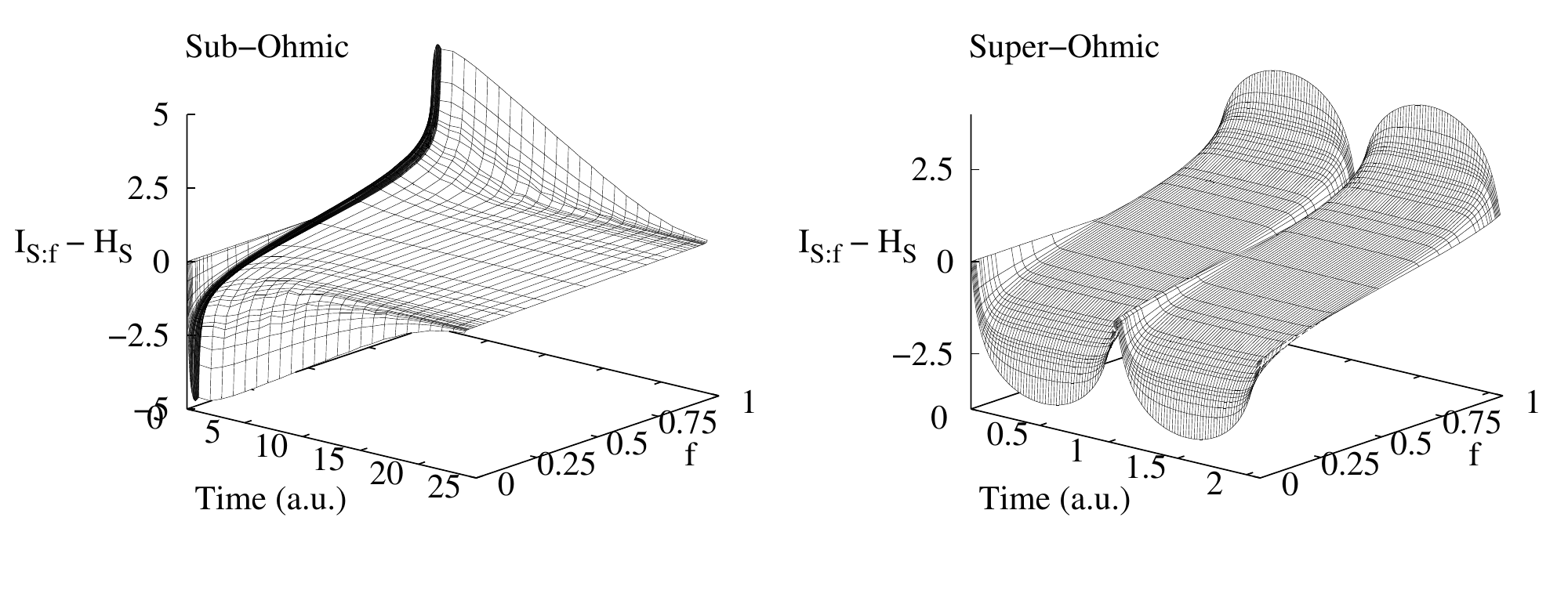}
\caption{Partial Information Plots (PI-Plots) for a dissipative (sub-Ohmic) environment (left) and 
non-dissipative (super-Ohmic) one (right). Initial state is squeezed in momentum with $r=-5$. 
The super-Ohmic case displays clear recoherence events for which the system and the environment loose their correlations. } 
\label{fig:PIP}
\end{figure}

\textit{Correlations with fractions of the environment:} 
We use ``partial information plots'' (PI-Plots) and ``partial entanglement plots" (PE-Plots) to study the evolution of
correlations between $\sys$ and fractions $\env_f$ of variable size $f$. They are plots of the average mutual information
(average of MI over random choices of fractions $\env_f$ of the same size) and average entanglement between the $\sys$ and $\env_f$,
respectively.
The evolution of the PI-Plots for different spectral densities are shown in Fig. \ref{fig:PIP}
(we verified that for the Ohmic and sub-Ohmic cases the results are similar), where for simplicity we subtract the
system's entropy and plot $\info(\sys,\env_f)-\entro(\sys)$. As the total state is pure, mutual information is 
$\info(\sys,\env_f)=\entro(\sys)+\entro(\sys,\env_{1-f}) -\entro(\sys,\env_f)$, i.e. it is symmetric around $f=1/2$
\cite{Blume}. This means that by adding the mutual information between the system and two complementary fractions of the
environment, one always obtains the maximal available information $\info(\sys,\env)=2\entro(\sys)$ 
(i.e. $\info(\sys,\env_f)+\info(\sys,\env_{1-f})=2\entro(\sys)$). Redundancy clearly manifests in 
the PI-Plots: There is a sharp growth of the mutual information for small $f$ followed by a plateau
that indicates that mutual information becomes almost independent of the fraction $f$. 

As PI-Plots  
are not completely flat around $f=1/2$ there is also some non-redundant information $\info_{NR}$ \cite{Blume} 
that can be defined as the slope of ${\info}(f)$ at $f=1/2$ (see below). Our results show that for dissipative
environments (both Ohmic and sub-Ohmic) $\info_{NR}$ is dramatically reduced, much faster than the total available
information. The behavior of the super-Ohmic environment is rather different: redundancy develops but this effect
is reversible. Decoherence is followed by recoherence and de-correlation.  Thus, the super-Ohmic environment does not
induce truly dissipative effects but is responsible for a renormalization (dressing) of the system as well as for 
coherent oscillations which will also be visible in our forthcoming studies. Perfect reversion for the super-Ohmic
environment only takes place for some initial states of the system. If instead of the above state we chose one 
with a large momentum squeezing complete recoherence is lost. This was already noticed in
\cite{PazHabZur93} and is due to transitory effects due to the instantaneous switching of the interaction. 

\begin{widetext}

\begin{figure}[ht]
\centering
\includegraphics[width=17cm]{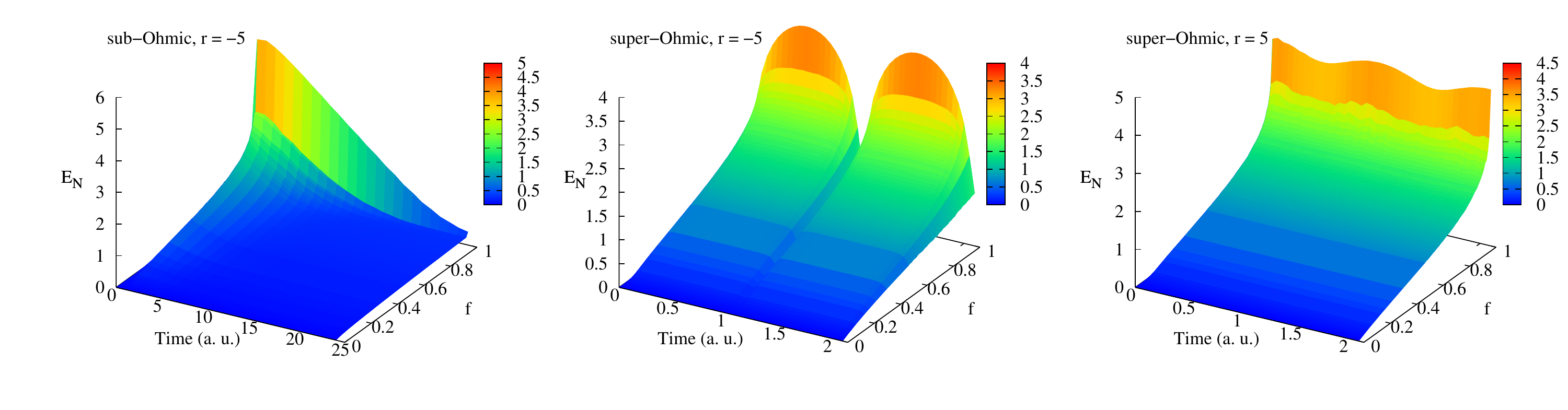}
\caption{Partial Entanglement Plot (PE-Plot) for sub-Ohmic (left) and super-Ohmic environments (center and right). 
The initial state of the system is squeezed in position with $r=-5$ except for the right plot where the squeezing 
is in momentum $r=5$. In that case recoherence is suppressed. Entanglement grows very slowly as a function of 
the fraction size $f$, which is evidence of the branch structure of the total state.} 
\label{fig:PEP}
\end{figure}
\end{widetext}

Partial entanglement plots (PE-Plots) in Fig. \ref{fig:PEP} show that for small
(and moderate) values of $f$, entanglement scales almost linearly with $f$ and stays very small. 
For larger fractions the rapid growth of entanglement indicates that it is established
between the system and the environment "as a whole". The effect of dissipation is displayed in
the PE-Plots: Dissipation reduces the maximum attainable entanglement and flattens the initial
slope of the PE-Plot, suppressing even more the quantum correlations with small
fractions of the environment. Super-Ohmic PE-Plot is dramatically different from the one corresponding 
to the dissipative case. In this non-dissipative environment revivals are evident in the PE-Plot. 
Redundancy of quantum correlations is related with the flatness of the PE-Plot as 
it indicates that many small environmental pieces become entangled as a whole with the system $\sys$. 
A way to quantify this notion of redundancy will be presented below.  

\section{Analytical results}
\label{sec:analyt}

\begin{figure}[ht]
\begin{flushleft}
\includegraphics[width=9cm]{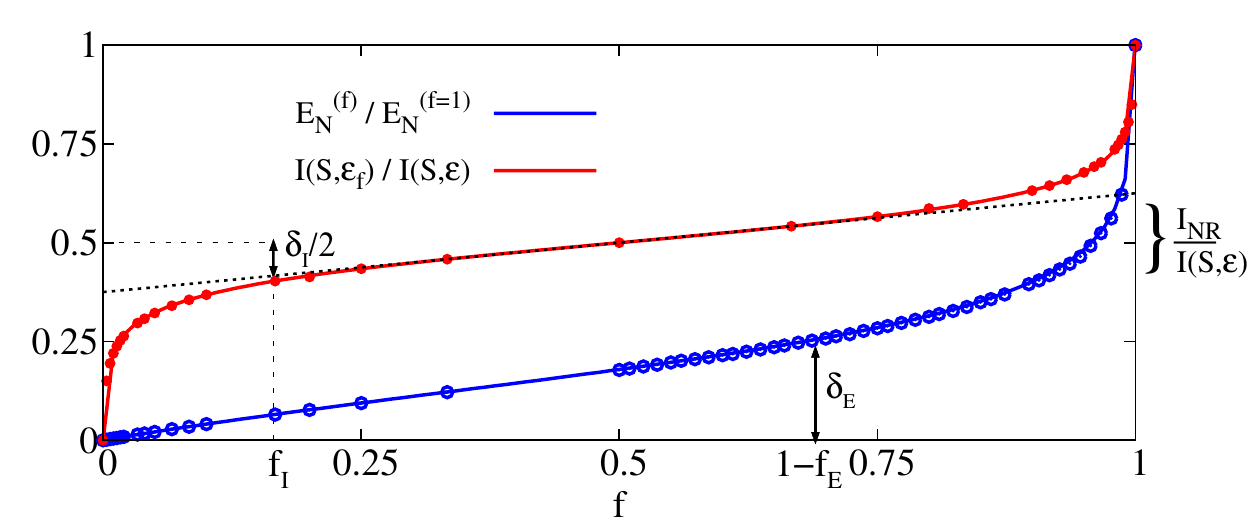}\\
\includegraphics[width=8.5cm]{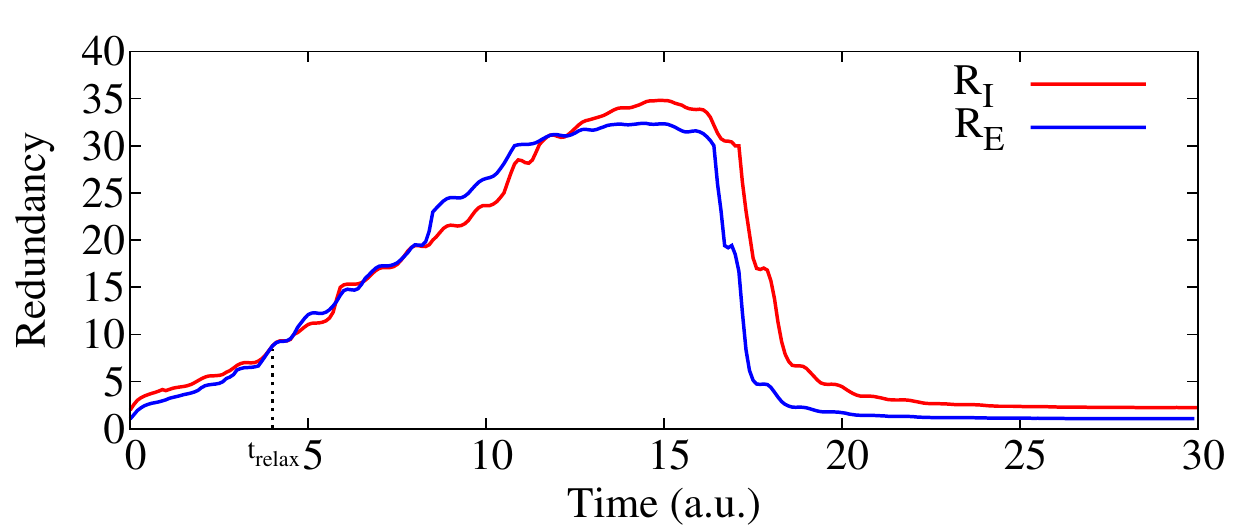}
\end{flushleft}
\caption{
\emph{Top:} Normalized PE-Plot and PI-Plot for a fixed time. Analytical estimates (solid lines) given by expressions in
equations (\ref{eq:entf}) and (\ref{eq:info}) show a remarkable agreement with numerical data (dots). 
The relevant fractions $f_E$ and $f_I$ to
compute information and entanglement redundancies $R_E$ and $R_I$ are shown in the figure. 
\emph{Bottom:} $R_E$ and $R_I$ as a function of time are shown for a dissipative environment ($n=1/2$), 
where the relaxation timescale is $t_{relax}\approx 3$ (with a deficit $\delta_E=2\delta_I=0.2$). 
} 
\label{fig:pepTheo}
\end{figure}

A simple model enables us to reproduce and understand the numerical results:
We assume that $\sys$ is massive and under-damped, justifying a Born-Oppenheimer approximation \cite{Blume08}. 
Under this approximation, the evolution of the joint state of $\sys-\env$ has a branch structure that can be written as:
\beq
|\Psi\rangle_{\sys\env}=\int\psi(x,t)\ket{x}_\sys\otimes\ket{\psi_x(t)}_{\env_1}\dots\otimes\ket{\psi_x(t)}_{\env_N}
{\rm d}x,
\label{eq:branch}
\eeq
where for each environmental oscillator, the state $\ket{\psi_x(t)}_{\env_n}$ depends parametrically on the system's
position $x$. Its time evolution can be found by noticing that each oscillator feels an external force $F_n(t)=c_n x(t)$.
Here, $x(t)$ parametrically depends on $x$ and has a rather simple form for large squeezings: For positive (negative) $r$
a trajectory starts from $x(0)=x$ ($x(0)=0$) with velocity $\dot x(0)=0$ ($\dot x(0)=\Omega_\sys x$). Thus, in general,
$x(t)=\theta(r) x \cos(\Omega_\sys t) + \theta(-r) x \sin(\Omega_\sys t)$. Using this, the fate of initial 
squeezed Gaussian states can be found by computing all elements of the covariance matrix of the full state. The 
entanglement $\ent^{(f)}$ can be analytically computed and turns out to depend upon the function  
$d^{(\env_f)}(t)=\sum_{n\in\env_f} d_n=\sum_{n\in\env_f}c_n^2\left({w^2_n a^2_n(t)}+ {\dot a}_n^2(t) \right)/4 m_nw_n$,
where 
\beqa
a_n(t)&=&{{\theta (r)}\over (w_n^2-\Omega^2_\sys)}
\left({\Omega_\sys\over w_n}\sin(w_n t)-\sin(\Omega_\sys t)\right)\quad \nonumber \\
&+&{{\theta (-r})\over (w_n^2-\Omega^2_\sys)}
\left(\cos(\Omega_\sys t)-\cos(w_n t)\right). \nonumber
\eeqa
On average, this satisfies $d^{(\env_f)}(t) =f \ d(t)$ and
\beqa
&&\ent^{(f)}(t)=-{1\over 2}\ln\Big\{1+4 d(t)\delta x^2 (1+3f) \label{eq:entf}\\
&&-4\sqrt{\left({f\over 1+3f}+d(t)\delta x^2(1+3f)\right)^2
-\left({f\over 1+3f}\right)^2}\Big\}.\nonumber 
\eeqa
where $\delta x$ is defined in terms of the absolute value of the squeezing parameter 
$|r|=\ln(\delta x/\delta x_0)$ being $\delta x_0$ the ground state variance $\delta
x_0=(1/2m\Omega_\sys)^{1/2}$. This result reproduces with high accuracy the numerical data before dissipation sets in,
see Fig. \ref{fig:pepTheo}; and replacing $f\rightarrow d_n(t)/d(t)$ we obtain the entanglement with the band $w_n$.
Before analyzing this, we present a similar result for the mutual information \cite{Blume08},
that turns out to be 
\beq
\info(\sys,\env_f)=h(\chi(1))+h(\chi(f))-h(\chi(1-f)). 
\label{eq:info}
\eeq
Here $h(\chi)$ is the entropy function $h(\chi)=(\chi+1/2)\ln(\chi+1/2)-(\chi-1/2)\ln(\chi-1/2)$ and its
argument $\chi(f)$ satisfies
\beq
\chi(f)=\sqrt{{1\over 4}+2 f\ d(t) \delta x^2}.
\eeq
The three terms in the equation (\ref{eq:info}) are, respectively,
the entropies for the system, the environment and the joint $\sys-\env_f$ ensemble 
(notice that $\chi(f=1)$ is the symplectic area of $\sys$).  
It is worth noticing that this simple model enables us to explain the peculiar super-Ohmic case (with high cutoff)
for which $d(t)$ is
\beq
d(t)\approx{m\gamma_0\over 2 \pi} \left(\theta(-r)\sin^2\Omega t
+\theta(r) (1+\cos^2\Omega t)\right).
\eeq
We can use this in the above formulae for $\ent^{(f)}$ and see that a dramatically different behavior for positive and
negative squeezing arises, fully reproducing the numerical data. For initial states delocalized in position the sudden
switch of the interaction produces a kick for each environmental oscillator whose effect cannot be undone later (the
environmental oscillators remain out of phase and no recoherence is possible). On the other hand, for initial states
delocalized in momentum the initial kick is negligible and the transient effects are minimized. An equivalent behavior is
observed if the interaction is adiabatically switched on. The existence of recoherence is not a
novel result \cite{PazHabZur93}, but is remarkable that it shows up in such a clear way for the entanglement 
and the mutual information.

$\ent^{(f)}$ and $\info(\sys,\env_f)$ have similarities but also remarkable differences. 
$\ent^{(f)}$ grows slowly for small
$f$ and fast for $f \approx 1$.  For $d(t)\delta x^2\gg1$, we have 
\beq
\ent^{(f)}\approx{1\over 2}\ln\left((1+3f)^3\over(1-f)(1+3f)^2 +2f/d(t)\delta x^2\right).
\eeq 
Numerical results are compared with the analytic estimate of $\ent^{(f)}$ in Fig. \ref{fig:pepTheo}. Notably, in the limit
of large squeezing and  $f<1$ the entanglement $\ent^{(f)}$ becomes independent of the environmental spectral density that
enters the above expressions through the function $d(t)$. In fact, in such case we have
\beq
\ent^{(f<1)}\approx{1\over 2}\ln\left(1+3f\over 1-f\right),
\eeq
a similar behavior was observed for 
Gaussian Greenberger-Horne-Zeilinger (GHZ)-type states \cite{Adesso04}. In the presence of dissipation $\ent^{(f)}$ is
reduced so the above universal expression is really an upper bound for $\ent^{(f)}$. A remarkably similar result arises 
for the mutual information. The analysis in this case is slightly different
since MI depends on the squeezing through the combinations $fd(t)\delta x^2$ and  $(1-f) d(t)\delta x^2$. Thus, both for
small and large fractions the first derivative of
$\info(\sys,\env_f)$ is singular (due to the singularity in $h(\chi)$). In the limit $d(t)\delta x^2\gg 1$,
$\entro(\sys)\approx\ln(2e^2d(t)\delta x^2)/2$ and mutual information behaves as  
\beq
\info(\sys,\env_{0<f<1})\approx \entro(\sys)+{1\over 2}\ln\left({f\over 1-f}\right). 
\label{eq:mi}
\eeq
$\info(\sys,\env_f)$ grows fast for small $f$ and approaches a plateau with a slope related to the so-called non-redundant
information $\info_{NR}$. This also turns out to be universal since  
$\info_{NR}\equiv{\partial{\info(\sys,\env_f)}/\partial f}|_{f=1/2}\approx 2$, see Fig. \ref{fig:pepTheo}. 

\section{Redundancy of correlations}
\label{sec:red}

Our results for $\ent^{(f)}$ enable us to define a new measure quantifying the redundancy of quantum correlations. We
define the \textit{entanglement redundancy}, $R_E=1/f_E$, as the number of environmental fractions of size $f_E$ such that by
ignoring one of them we induce a decay of the average entanglement to a fraction $\delta_E$ of the maximal one. Roughly speaking, a
large entanglement redundancy is obtained when by ignoring a small fraction of the environment one looses a large fraction
of the total entanglement. This is the case if $f_E$ turns out to be small when $\delta_E$ is small. 
GHZ states have large entanglement redundancy, since by tracing over one of its constituents the entanglement is lost.  
Indeed, large entanglement redundancy characterizes the fragility of the multipartite entanglement. 
Computing $R_E$ from the PE-Plots is simple: As shown in
Fig. \ref{fig:pepTheo}, we should find the smallest fraction $f_E$ such that 
$\ent^{(1-f_E)}=\delta_E\ent^{(f=1)}$. Since $R_E$ is the number of such
fractions in the environment $\env$, we have $R_E=1/f_E$. As $\ent^{(f)}$ is a monotonous function of $f$, a large
redundancy is expected for branch states eq. (\ref{eq:branch}), in that case $\ent^{(f)}$ would grow very slowly with $f$ for small and
moderate values of $f$. Remarkably, our analytic results before dissipation enable us to obtain a good estimate for $R_E$: 
\beq
R_E={1\over{f_E}}\approx e^{4\delta_E\ent^{(f=1)}}\approx A(t)^{2\delta_E},
\eeq
where $A(t)= d(t)\delta x^2/\delta_0 x^2$ is the symplectic area of the system's state 
(in the limit of large squeezings) and $\ent^{(f=1)}\approx\ln(32d(t)\delta x^2)/2$ is the 
maximal entanglement (which is proportional to the entropy of the
system). 

It is interesting to notice that entanglement redundancy $R_E$ can be related with \textit{information redundancy} $R_I$,
defined in \cite{Blume08} as the number of environmental pieces, $R_I=1/f_I$, that carry a fraction $(1-\delta_I)$ of the available
classical information $\entro(\sys)$ (see Fig. \ref{fig:pepTheo}).  
In Fig. \ref{fig:pepTheo}, we can see that both redundancy measures, $R_E$ and $R_I$, behave in a similar way. 
Notably, both $R_E$ and $R_I$ grow for times much longer than the relaxation timescale. 
This is a consequence of the fact that when dissipation becomes effective the quantum correlations with fractions of the
environment are almost erased. Consequently, PE--Plots and PI--Plots flatten well before equilibration. 
Finally, in the asymptotic limit, redundancy almost disappears as do all types of correlations when the
system approaches equilibrium. Our results, also show that the amount of non-redundant information is
intimately related with the entanglement created between $\sys$ and fractions of the environment. Besides, we can show
that both redundancies numerically coincide when the deficits $\delta_E$ and $\delta_I$ are such that 
$\delta_E\approx \delta_I \entro(\sys)/\ent^{(f=1)}+\ent^{(f=1/2)}/\ent^{(f=1)}$. Thus, in the limit of large squeezing
both deficits become identical (this is the case because $\ent^{(f=1/2)}$ is bounded from above as 
$\ent^{(f=1/2)}\le \ln\sqrt 5$).

\section{Conclusions}
\label{sec:conc}

In this paper we studied the evolution of quantum correlations (characterized by the entanglement) and total correlations (measured by the MI) between a system and its environment. First we analyzed the evolution of such correlations as a function of the band's frequency. For this case we showed that for dissipative
environments the resonant bands are always the ones that correlate the most with the system. On the other hand, for the non-dissipative environments the high frequency bands dominate the evolution inducing short time effects that can leave a long time footprint for some initial states. Other initial states lead to revivals of the correlations. Our analytic and numerical study of the evolution of correlations for QBM shows that MI and entanglement do become redundant at a time-scale which is connected with the one of decoherence
(indeed, the function $d(t)$ appearing above is the same that characterizes the exponential suppression of the quantum interference effects in QBM). This is seen in the PI-Plots and PE-Plots. 
Our results also show the special nature of the state that dynamically evolve for the QBM model that has a 
branch structure compatible with large entanglement redundancy and truly multipartite entanglement 
between $\sys$ and $\env$. Dissipation rapidly erases quantum correlations with fractions of the environment
and flattens both PI-Plots and PE-Plots. Thus, redundancy is dramatically increased by the action of dissipation.

\acknowledgments
JPP is a member of CONICET. Authors acknowledge support of CONICET, UBACYT and ANPCyT (Argentina).

\end{document}